\documentclass{elsart}

\usepackage{harvard}

\usepackage{epsfig}

\def\url#1{{\ttfamily\def\/{/\discretionary{}{}{}}#1}}

\def\simlt{\hbox{ \rlap{\raise 0.425ex\hbox{$<$}}\lower 0.65ex\hbox{$\sim$} }}
\def\simgt{\hbox{ \rlap{\raise 0.425ex\hbox{$>$}}\lower 0.65ex\hbox{$\sim$} }}

\begin{document}

\begin{frontmatter}
\title{An efficient parallel algorithm for $O(N^2)$ direct summation
method and its variations on distributed-memory
parallel machines}

\author{Junichiro Makino}

\address{Department of  Astronomy,\\
School of Science, University of Tokyo,\\
7-3-1 Hongo, Bunkyo-ku, Tokyo 113-0033, Japan.
}

\def\titleabbrev{{Parallelization of Individual timestep}}
\def\undertext#1{{$\underline{\hbox{#1}}$}}
\def\doubleundertext#1{{$\underline{\underline{\hbox{#1}}}$}}
\def\half{{\scriptstyle {1 \over 2}}}
\def\ie{{\it {\frenchspacing i.{\thinspace}e. }}}
\def\eg{{\frenchspacing e.{\thinspace}g. }}
\def\cf{{\frenchspacing\it cf. }}
\def\etal{{\frenchspacing\it et al.}}
\def\et{{\etal}}
\def\simlt{\hbox{ \rlap{\raise 0.425ex\hbox{$<$}}\lower 0.65ex\hbox{$\sim$} }}
\def\simgt{\hbox{ \rlap{\raise 0.425ex\hbox{$>$}}\lower 0.65ex\hbox{$\sim$} }}
\def\solar{\odot}
\def\msun{\ifmmode{M_\solar}\else{$M_\solar$}\fi}
\def\rsun{\ifmmode{R_\solar}\else{$R_\solar$}\fi}
\def\Rf{\parindent=0pt\smallskip\hangindent=3pc\hangafter=1}
\def\pc{{\rm pc}}
\def\kpc{{\rm kpc}}
\def\Mpc{{\rm Mpc}}
\def\yr{{\rm yr}}
\def\Myr{{\rm Myr}}
\def\Gyr{{\rm Gyr}}
\def\kT{\ifmmode{kT}\else{$kT$}\fi}
\def\N{{\ifmmode{N}\else{$N$}\fi}}
\def\fb{{\ifmmode{f_B}\else{$f_B$}\fi}}
\def\emax{{\ifmmode{E_{max}}\else{$E_{max}$}\fi}}
\def\td{{\ifmmode{t_d}\else{$t_d$}\fi}}
\def\tcr{{\ifmmode{t_{cr}}\else{$t_{cr}$}\fi}}
\def\tr{{\ifmmode{t_r}\else{$t_r$}\fi}}
\def\trh{\ifmmode{t_{rh}}\else{$t_{rh}$}\fi}
\def\vv{{\ifmmode{\langle v^2\rangle}\else{$\langle v^2 \rangle$}\fi}}
\def\v{{\ifmmode{\langle v^2\rangle^{1/2}}
		\else{$\langle v^2 \rangle^{1/2}$}\fi}}
\def\half{{\ifmmode{{1 \over 2}}\else{${1 \over 2}$}\fi}}
\def\dhalf{{\textstyle {1 \over 2}}}
\def\threehalf{{\ifmmode{{3 \over 2}}\else{${3 \over 2}$}\fi}}
\def\dthreehalf{{\textstyle {3 \over 2}}}
\def\dfivehalf{{\textstyle {5 \over 2}}}
\def\dfivethree{{\textstyle {5 \over 3}}}
\def\kms{\ifmmode{\rm km\,s^{-1}}\else{$\rm km\,s^{-1}$}\fi}
\def\kmps{{\rm km/s}}
\def\piet#1{{\bf[#1 -- piet]}}
\def\jun#1{{\bf[#1 -- jun]}}
\def\bx{{\bf x}}
\def\bv{{\bf v}}
\def\ba{{\bf a}}
\def\badot{{\bf \dot a}}
\def\batwo{{{\bf  a}^{(2)}}}
\def\bathree{{{\bf  a}^{(3)}}}
%
%
\font\lgh=cmbx10 scaled \magstep2

\begin{abstract}

We present a novel, highly efficient algorithm to parallelize
$O(N^2)$direct summation method for $N$-body problems with
individual timesteps  on distributed-memory parallel machines
such as Beowulf clusters. Previously known algorithms, in which all
processors have  complete copies of the $N$-body system, has the
serious problem that the communication-computation ratio increases as
we increase the number of processors, since the communication cost is
independent of the number of processors. In the new algorithm,
$p$ processors are organized as a $\sqrt{p}\times
\sqrt{p}$ two-dimensional array. Each processor has  $N/\sqrt{p}$
particles, but the data are distributed in such a way that complete
system is presented if we look at any row or column consisting of
$\sqrt{p}$ processors.  In this algorithm, the communication cost scales as
$N /\sqrt{p}$, while the calculation cost scales as
$N^2/p$. Thus, we can use a much larger number of processors without
losing efficiency compared to what was practical with previously known
algorithms.

{\it PACS: 02.60.Cb;95.10.Ce; 98.10.+z} 

\begin{keyword}
Celestial mechanics, stellar dynamics;Methods: numerical
\end{keyword}

\end{abstract}

\end{frontmatter}

\section{Introduction}

In this paper we present a novel algorithm to parallelize the direct
summation method for astrophysical $N$-body problems, either with and
without the individual timestep algorithm. The proposed algorithm
works also with the Ahmad-Cohen neighbor scheme
\cite{AhmadCohen1973}, or with GRAPE special-purpose computers for
$N$-body problems \cite{Sugimoto1990,MakinoTaiji1998}.  Our algorithm
is designed to offer better scaling of the communication-computation
ratio on distributed-memory multicomputers such as Beowulf PC clusters
\cite{Sterlingetal1999} compared to traditional algorithms.

This paper will be organized as follows. In section 2 we describe the
traditional algorithms to parallelize direct summation method on
distributed-memory parallel computers, and the scaling of
communication time and computational time as functions of the number
of particles $N$ and number of processor $p$. It will be shown that
for previously known algorithms the calculation time scales as
$O(N^2/p)$, while communication time is $O(N+\log p)$. Thus, even with
infinite number of processors the total time per timestep is still
$O(N)$, and we cannot use more than $O(N)$ processors without losing
efficiency. $O(N)$ sounds large, but the coefficient is rather
small. Thus, it was not practical to use more than 10 processors for
systems with a few thousand particles, on typical Beowulf clusters.  

In section 3 we describe the basic idea of our new algorithm. It will
be shown that in this algorithm the communication time is
$O(N/\sqrt{p}+\log {p})$. Thus, we can use $O(N^2)$ processors without losing
efficiency. This implies a  large gain in speed for relatively small
number of particles such as a few thousands.  We also briefly discuss
the relation between our new algorithm and the hyper-systolic
algorithm \cite{Lippertetal1998}. In short, though the ideas
behind the two algorithms are very different, the actual communication 
patterns are quite similar, and therefore the performance is also
similar for the two algorithms. Our algorithm shows a better scaling
and also is much easier to extend to individual timestep and
Ahmad-Cohen schemes. 

In section 4 we discuss the combination of our proposed algorithm and
individual timestep algorithm and the Ahmad-Cohen scheme. In section 5, we
present examples of estimated performance. In section 6 we discuss the 
combination of our algorithm with GRAPE hardwares. In
section 7 we sum up.

\section{Traditional approaches}

The parallelization of the direct method has been regarded simple and
straightforward [see, for example, \cite{PCW}]. However, it is only so
if $N>>p$ and if we use simple shared-timestep method. In this
section, we first discuss the communication-calculation ratio of
previously known algorithms for the shared timestep method, and then
those for individual timestep algorithm with and without the Ahmad-Cohen
scheme.

\subsection{Shared timestep}

\begin{figure}
\begin{center}
\leavevmode
\epsfxsize = 8 cm
\epsffile{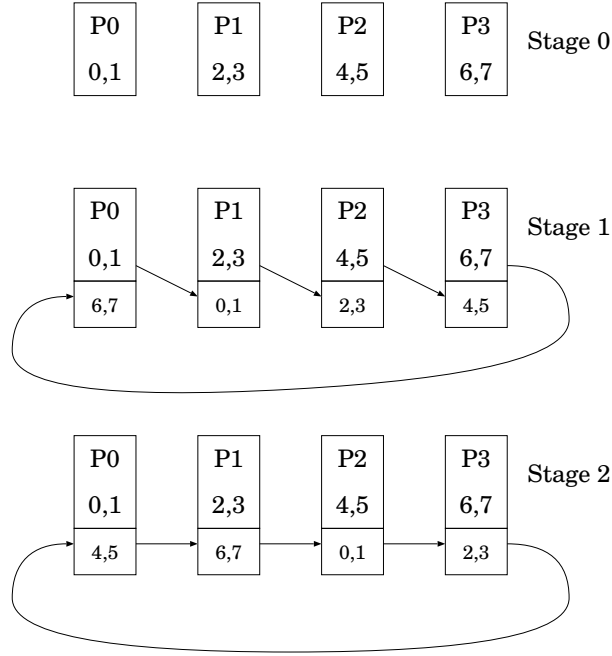}
\caption{The ring algorithm on 4 processors for 8 particles. In stage
0 each processor calculate the interactions between its particles. In stage 1 
each processor sends its share to its right neighbor, and calculates
the force from particles it received to particles it has. In stages 2 and 3 
(stage 3 omitted from the figure), each processor sends what it
received in the previous stage. Force calculation completes at stage 3.}
\label{fig:ring}
\end{center}
\end{figure}

Most of the textbooks and papers discuss the ring algorithm. Suppose
we calculate the force on $N$ particles using $p$ processors. We
connect the processors in a one dimensional ring, and distribute $N$
particles so that each processor has $N/p$ particles(figure
\ref{fig:ring}). Here and hereafter, we assume that $N$ is integer
multiple of $p$, to simplify the discussion.

The ring algorithm calculates the forces on $N$ particles in the
following steps.

\begin{enumerate}
\item Each processor calculates the interactions between $N/p$
particles within it. Calculation cost of this step is $C_f(N/p)^2/2$,
where $C_f$ is the time to calculate interaction between one pair of
particles.

\item Each processor sends all of its particles to the same
direction. Here we call that direction ``right''. Thus all processors
sends its particles to their right neighbors. The communication cost is
$C_cN/p + C_s$, where $C_c$ is the time to send one particle to the
neighboring processor and $C_s$ is the startup time for communication.

\item Each processor accumulates the force from particles they
received to its own particles. Calculation cost is $C_f(N/p)^2$. If force from all 
particles is accumulated, go to step 5.

\item Each processor then sends the particles it received in the
previous step to its right neighbor, and goes back to previous step.

\item Force calculation completed.

\end{enumerate}

The time for actual  calculation is given by
\begin{equation}
T_{\rm f,ring} = C_f N^2/p,
\end{equation}
and the communication time
\begin{equation}
T_{\rm c,ring}=C_cN + C_sp.
\end{equation}

The total time per one timestep of this algorithm is
\begin{equation}
T_{\rm ring} = T_{\rm f,ring} + T_{\rm c,ring}=C_f N^2/p + C_cN + C_sp.
\label{eq:tring}
\end{equation}
Here, we neglect small correction factors of order $O(1/p)$. 

For fixed number of particles, the calculation cost (first term in
equation \ref{eq:tring}) scales as $1/p$ while communication cost {\it 
increases}. Therefore, for large $p$ we see the decrease in the
efficiency. Here we define efficiency as
\begin{equation}
\eta_{\rm ring} = T_{\rm f,ring} /T_{\rm ring},
\end{equation}
which reduces to
\begin{equation}
\eta_{\rm ring} = \frac{1}{1 + C_cp/(C_fN) + C_sp^2/(C_fN^2)}.
\end{equation}
Thus, to achieve the efficiency better than 50\%, the number of
processor $p$ must be smaller than
\begin{equation}
p_{\rm half,ring} = \frac{N}{2C_s}(\sqrt{C_c^2+4C_s C_f} -C_c).
\label{eq:phalfring}
\end{equation}

Equation (\ref{eq:phalfring}) can be simplified in the two limiting
cases
\begin{equation}
p_{\rm half,ring} \sim \cases{NC_f/C_c & ($C_c^2 >>C_s C_f$)\cr
                              N\sqrt{C_f/C_s} & ($C_c^2 <<C_s C_f$)
}
\end{equation}
In most of distributed-memory multicomputers, $C_c >> C_f$. For
example, with a 1 Gflops processor, we have $C_f \sim 3\times
10^{-8}{\rm sec}$. If this processor is connected to other processor
with the communication link of the effective speed of 10MB/s, $C_c
\sim 3\times 10^{-6}{\rm sec}$.  The value of $C_s$ varies depending
on both networking hardware and software. Table 1 gives the
order-of-magnitude values for these coefficients for several
platforms.

\begin{table}
\caption{Time coefficients in seconds}
\begin{tabular}{lccccc}
\hline
Network & $C_f$ & $C_s$ & $C_c$ & $\sqrt{C_f/C_s}$ & $C_f/C_c$\\
\hline
Fast Ethernet &$3\times 10^{-8}$ & $ 10^{-4}$ & $ 3\times 10^{-6}$ & 0.017 & 0.01\\
Gigabit Ethernet &$3\times 10^{-8}$ & $ 10^{-4}$ & $ 3\times 10^{-7}$ & 0.017 & 0.1\\
Myrinet &$3\times 10^{-8}$ & $ 10^{-5}$ & $ 3\times 10^{-7}$ & 0.05 &
0.1\\
\hline
\end{tabular}
\end{table}
With the Fast Ethernet, $p_{\rm half,ring}$ is $N/100$. Faster
networks help to increase $p_{\rm half,ring}$, but we can see that even
with a very low-latency network system like Myrinet, $p_{\rm half,ring}$
is latency-limited. Fast, high-latency network such as Gigabit
Ethernet does not offer much improvement over Fast Ethernet.

\begin{figure}
\begin{center}
\leavevmode
\epsfxsize = 8 cm
\epsffile{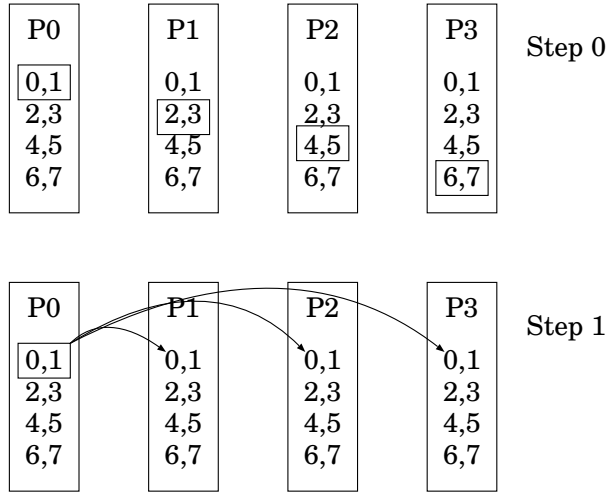}
\caption{The copy algorithm on 4 processors for 8 particles. In stage 0 
each processor calculates the forces on its share of particles and
integrates their orbits. In stage 1, processor 0 broadcasts its share
of particles to all other processors. In the following stages (not
shown in the figure), each processor broadcasts its share in turn.}
\label{fig:copy}
\end{center}
\end{figure}

A possible alternative of the ring algorithm is the copy algorithm
(figure \ref{fig:copy}), in which all processors have complete copy of
the system at the beginning of the timestep. Then, each processor
calculates the forces on its share of $N/p$ particles, and integrates
their orbits. After the orbit integration is done, the updated
particles in each processor must be broadcasted to all other
processors. If we use a simple ring algorithm to broadcast the data,
the scaling characteristic of this algorithm is exactly the same as
that of the ring algorithm.

If the communication network has the connectivity better than 1-D
ring, we can use the so-called message-combining technique, which is
essentially the same algorithm as what is used for global summation.
Assume that we have $2^s$ processors with fully connected network. At
the first stage, each processor sends its data to its right
neighbor. At the second stage, each processor sends  both its original
data and the data it received at the first stage to its second right
neighbor, and in the third stage to fourth neighbor. In this way, all 
processors receive the data of all other processors after $s = \log_2
p$ stages. Note that the amount of the data received is the same as
the ring algorithm. Only the startup overhead is reduced. With this
implementation of the copy algorithm, the total time becomes
\begin{equation}
T_{\rm copy} = T_{\rm f,ring} + T_{\rm c,ring}=C_f N^2/p + C_cN +
C_s\log_2 p.
\label{eq:tcopy}
\end{equation}
In this case, we can almost always ignore the contribution of the last 
term, and the number of processors with 50\% efficiency is given by
\begin{equation}
p_{\rm half,copy} \sim NC_f/C_c. 
\end{equation}

Thus, we can use a larger number of processors, as far as the memory of
individual node is large enough to keep the copy of the complete
system. We could also implement some hybrid of the ring and copy
algorithm to reduce the memory requirement. 

\subsection{Individual timestep}

Here, what we actually consider is the so-called blockstep
method \cite{McMillan1986,Makino1991a}, where we organize the timesteps
of particles so that the number of particles share exactly the
same time is becomes maximum. This is clearly desirable to achieve high efficiency on
parallel computers. We denote the average number of particles which
share the same time as $n$. According to a simple theoretical estimate,
$n\propto N^{2/3}$ if the central density of the system is not very
high \cite{Makino1991a}.

It's not easy to extend the ring algorithm so that it can
handle the blockstep algorithm. A naive approach is just to pass
around all particles in the system in the same way as in the case of
the shared timestep. In this case, calculation cost is $C_fNn/p$ while 
communication cost is $C_cN$. Thus, relatively speaking communication
becomes more expensive by a factor of $N/n$, which is essentially the
gain of calculation speed by the individual timestep algorithm.

The copy algorithm is much easier to extend to the blockstep
method. At each blockstep, each processor determines which particles
it will update, so that it updates $n/p$ particles which are not
overlapped with particles updated by other processors.  Then they
calculate the forces on them and integrate their orbits. At the end of
the timestep, they broadcast the updated particles. The calculation
cost is $C_fNn/p$, while the communication cost is $C_cn+C_s\log_2p$. Thus, the
communication is reduced by the same factor as that for the calculation cost.

Note, however, that the effect of the startup overhead of
communication is increased. We can still 	
ignore the  startup time  if
\begin{equation}
\log_2 p C_s < C_c n,
\label{eq:indcrit}
\end{equation}
which is okay except for the case of the Gigabit ethernet. 
Thus, individual timestep does not change the scalability of the algorithm.

We cannot easily use more than $n$ processors, since each processor
should have at least one particles to integrate.  At least under the
assumption of $n = N^{2/3}$, $p_{\rm half,copy}$ becomes larger than
$n$ for $N> 1000$, for the case of Myrinet. In this case, we can
distribute the work to calculate the force on one particle to several
processors. This of course increases the communication cost, but the
increase is around $C_c p/n$ which is always smaller than $C_c n$.

In theory, it is possible to extend the ring algorithm in the
following way. Instead of passing around the particles which exert the 
forces, we can pass around the particles which receives the force. In
this way, the ring algorithm can achieve the same scaling as the copy
algorithm. The naive use of the ring communication pattern again
incurs the high startup cost. So it is necessary to use the message
combining technique. 

The ring algorithm suffers a potential load imbalance problem, since
particles are assigned to fixed processors. At any given blockstep,
there is no guarantee that the number of particles to be integrated is
balanced. Of course, we can try to migrate particles with small
timestep from heavily loaded processors to lightly loaded processors,
and vice versa, to reduce the imbalance. The effectiveness of such an
algorithm is an open question, though we expect the load balancing
would probably work fine.

For the individual timestep method, however, there is no reason to
prefer the (modified) ring algorithm over the copy algorithm, since
the efficiency of the ring algorithm at the best only matches that of
the copy algorithm. The potential advantage of the ring algorithm is
that it requires less memory, since the particles are distributed over
processors. However, since the number of particles we consider here is
relatively small, the memory requirement is not very important.

\subsection{The Ahmad-Cohen scheme}

The copy algorithm is problematic to extend to the Ahmad-Cohen scheme. The problem is
that informations to be passed around increases since now each
particle has its own neighbor list. This list must be communicated to
all processors. On the other hand, the calculation cost is further
reduced. 

The ring algorithm is better than the copy algorithm here, since with
the ring algorithm we do not have to pass around the neighbor
list. Each processor maintains partial neighbor lists for all
particles, which contains only the particles on that processor. The
scaling relation of calculation and communication costs are
essentially the same as that of the individual timestep, except that
now total number of particles $N$ is replaced with average number of
force calculations per timestep. Theoretically, it is of the order of
$N^{3/4}$\cite{MakinoHut1988}. This means we can now use only
$N^{3/4}{C_f/C_c}$ processors. Even with Myrinet, this number is
rather small. For example, we can use only 100 processors for
$N=10^4$.

\subsection{Summary}

We overviewed  known algorithms to parallelize the direct force
calculation. All algorithms share the same problem that the relative cost 
of the communication goes up as we increase the number of
processors. More troublesome is that even for fixed $N$ and  $p$, the
relative cost of communication is higher for a more advanced algorithm,
resulting in the partial cancelation of the gain in the speed
achieved by advanced algorithms.

We have not discussed the ``hyper-systolic''
algorithm\cite{Lippertetal1998}. We will briefly discuss them after we
introduce our new algorithm in the next section.

\section{The new algorithm}

The problem with the traditional schemes is that each processor must
receive the information of all particles updated at each
timestep. This is the case for all algorithms described in the previous
section. Therefore, the communication cost remains constant, while
calculation cost decreases as we increase $p$.

If all processors keep the complete copies of the system, it is clear
that they must receive all data updated by other processors at each
timestep. If all processors only store their own shares, they still have 
to receive all updated particles to calculate forces either to or from 
them.

If we divide the processors to subgroups, and let the data  be copied
only within the subgroups, we may be able to reduce the communication
cost. In the following, we'll discuss how such subdivision works.

Here, we assume that we have $p=r^2$ processors, where r is a positive
integer, and that they are organized as $r\times r$ 2D grid
(torus). processors are numbered as $q_{0,0}$ to $q_{r-1,r-1}$. It is
easy to extend the new algorithm to any rectangular grid with the geometry
of $r \times kr$, where k is a positive integer. Such configuration
can reduce the memory requirement, but increases the communication
cost. Therefore we do not discuss such non-square grids in this paper.

\begin{figure}
\begin{center}
\leavevmode
\epsfxsize = 8 cm
\epsffile{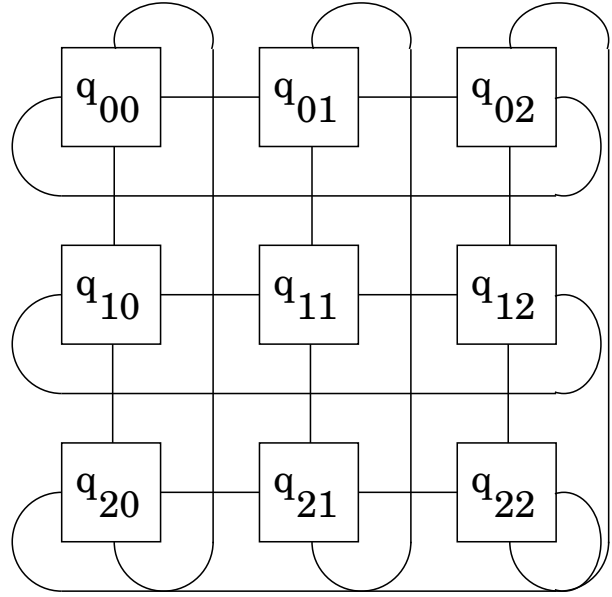}
\caption{Two dimensional array of processors. In this example, a torus 
network is shown.}
\label{fig:twodarray}
\end{center}
\end{figure}

We divide  $N$ particles to $r$ subsets each with $N/r$ particles, and 
let processors $q_{x, i}$ (here $x,i$ means  any index pair with
column address $i$) to have $i$-th subset. Thus, processors $q_{0,0}$
through $q_{r,0}$ have the same data of zeroth group.

The force calculation can be done by the following two steps. First, we 
apply the ring algorithm for each row. However, instead of sending all 
particles in each processor, they send only $N/r^2$ particles. To do
so, within processors in the same column the particles are further
subdivided to $r$ sub-subgroups.

After one rotation of the ring is finished, each processor obtained
the partial force from $N/r$ particles in the ring (row) to $N/r$
particles in the column. Calculation cost of this stage is $C_fN^2/p$
and communication cost is $C_cN/r + C_s r$, since we send $N/r^2$ particles
$r$ times.

In the second stage, we simply take the summation of $r$ partial
forces for each particles, which are distributed on the $r$ processors
in the same column.  The total force is obtained on one processor,
which then broadcasts them to all other processors in the same column.
The communication cost depends very much on the connectivity of the
network. In the worst case of 1-D network, the communication cost of
this stage is $C_cN$. The summation takes $r$ steps in 1-D ring, and
each processor has to send $N/r$ data. Of course, if we can overlap
calculation and communication, the scaling can be $2C_cN/r+C_sr$,
which is far more preferable.  With full crossbar networks in this
column direction, the communication cost becomes $2C_cN\log_2 r/r + 2C_s\log_2
r$, assuming that we cannot overlap communication and calculation. If
we can overlap them, the communication cost is reduced to
$2C_cN/r+2C_s\log_2 r$.  In the following, for simplicity we assume
that the communication cost of this stage is the same as that for the
first stage.

\begin{figure}
\begin{center}
\leavevmode
\epsfxsize = 8 cm
\epsffile{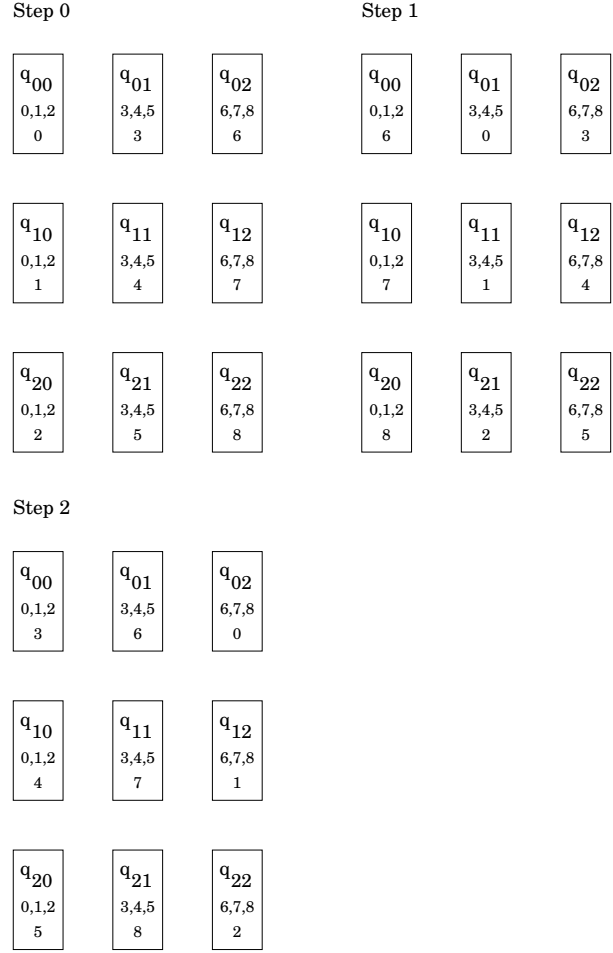}
\caption{The 2D ring algorithm.}
\label{fig:twodring}
\end{center}
\end{figure}

The total time per one timestep of this algorithm is
\begin{equation}
T_{\rm 2Dring} = T_{\rm f,2Dring} + T_{\rm c,2Dring}=C_f N^2/r^2 + 2(C_cN/r + C_sr).
\label{eq:t2D}
\end{equation}

For the number of processors $p=r^2$ for which the efficiency is 50\%, 
we have a cubic equation. For the two limiting cases, the solution is
given as

\begin{equation}
p_{\rm half,2Dring} \sim \cases{(NC_f/C_c)^2 & ($N<N_{\rm c,2Dring}$),\cr
                              N^2(C_f/C_s)^{2/3} & ($N>N_{\rm c,2Dring}$),
}
\label{eq:p2Dring}
\end{equation}
where $N_{\rm c,2Dring}$ is defined as
\begin{equation}
N_{\rm c,2Dring} = \frac{C_c^3}{C_f^2 C_s}
\end{equation}
The values of $N_{\rm c,2Dring}$ for typical networks are given in table \ref{tab:ncrits}.
Clearly, $N$ is almost always larger than $N_{\rm c,2Dring}$. This means
that the total  performance is limited by the latency of the network.

\begin{table}
\caption{Critical values of $N$ for different algorithms}
\label{tab:ncrits}
\begin{tabular}{lccc}
\hline
Network & $N_{\rm c,2Dring}$ & $N_{\rm c,2Dbcast}$ & $N_{\rm c,ind,2D}$ \\
\hline
Fast Ethernet & 300 & $3\times 10^5$ & $900$ \\
Gigabit Ethernet & 0.3 & $90$ & $200$ \\
Myrinet & 3 & 60  & 70\\
\hline
\end{tabular}
\end{table}

Even so, the number of processors we can use with this 2D algorithm is
significantly larger than that for 1D ring, for any value of $N$. If
$N<N_c$, we can use $O(N^2)$ processors. Even if $N>N_{\rm c,2Dring}$, we
can still use $O(N^{4/3})$ processors.

In this 2D ring algorithm,  the $O(r)$ term in the
communication cost limits the total performance. We can reduce this
term by using the extension of the copy algorithm to 2D.

\begin{figure}
\begin{center}
\leavevmode
\epsfxsize = 8 cm
\epsffile{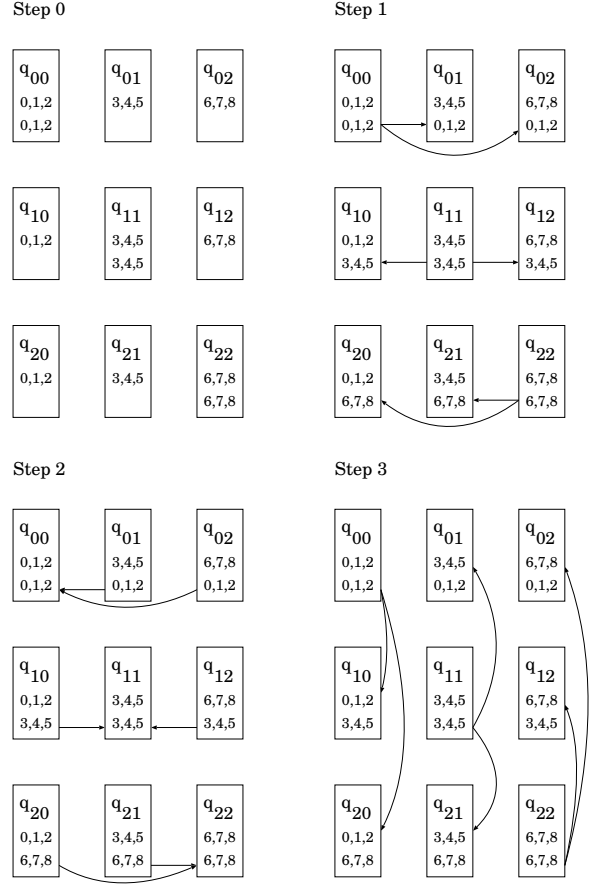}
\caption{The 2D broadcast algorithm.}
\label{fig:twodbroadcast}
\end{center}
\end{figure}

Instead of using the ring algorithm in the first stage, processors
$q_{ii}$ broadcast their data to all other processors in the same row. 
After this broadcast processor $q_{ij}$ has both group $i$ and group
$j$.  Then each processor calculates the force on particles they
received (group $i$) from particles they originally have (group $j$).
In this scheme, the communication cost
is reduced to $NC_c/r+C_s$ if the network switch supports the
broadcast.  If the network does  not support the broadcast, the cost
varies between $C_cN/r+C_sr$ for the case of a ring network and
$C_cN/r+C_s\log_2 r$ for a full crossbar.

In the second stage, summation is now taken over the processors in the
same row. Here, result for row $i$ must be obtained on processor
$q_{ii}$, which then broadcasts the forces to all processors in the
same column. After this broadcast, all processors have the forces on
all particles in them.  They can then use this forces to integrate the
orbits of particles.

In this algorithm, the time integration calculation is duplicated over
$r$ processors in the same column, but in most cases this does not
matter. One alternative is that processor $q_{ii}$ performs the time
integration and broadcasts the updated data of particles to other
processor in the same column. Yet another possibility is to let each
of $r$ processor to integrate $N/r^2$ particles, which each of them then
broadcasts within the column. Which approach is the best depends on the 
ratio between calculation speed, communication speed and communication 
startup overhead.

The total time per one timestep of this algorithm is
\begin{equation}
T_{\rm 2Dbcast} = T_{\rm f,2Dbcast} + T_{\rm c,2Dbcast}=C_f N^2/r^2 + 2(C_cN/r + C_s\log_2r).
\label{eq:t2Dbcast}
\end{equation}

For the number of processors $p=r^2$ for which the efficiency is 50\%, 
we have a cubic equation. For the two limiting cases, the solution is
given as

\begin{equation}
p_{\rm half,2Dbcast} \sim \cases{(NC_f/2C_c)^2 & ($N<N_{\rm c,2Dbcast}$),\cr
                              N^2C_f/[C_s\log 2\cdot (N^2C_f/2C_s)] & ($N>N_{\rm c,2Dbcast}$),
}
\label{eq:p2Dbcast}
\end{equation}
where $N_{\rm c,2Dbcast}$ is defined as
\begin{equation}
N_{\rm c,2Dbcast} = \sqrt{\frac{C_s}{C_f}}\exp\left(\log_2 \frac{2C_c^2}{C_fC_s}\right).
\end{equation}
The critical value of $N$, $N_{\rm c,2Dbcast}$, is larger than that
for the 2D ring version of the algorithm. This is because we reduced
the $C_sr$ term in the communication cost to $C_s\log r$. More
importantly, even for $N > N_{\rm c,2Dbcast}$, $p_{\rm half}$ is only
logarithmically smaller than $O(N^2)$.  Thus, with this broadcast
version of the algorithm we can really use $O(N^2)$ processors and
still achieve high efficiency.

\subsection{Relation with the hyper-systolic algorithm}

Now the relation between our algorithm and the hyper-systolic
algorithm \cite{Lippertetal1998} must be obvious. The ``regular
bases'' version of the hyper-systolic algorithm applied to $r^2$
processors works in the essentially the same way as the ring version
of our algorithm works, though in order to derive our algorithm we do
not need to use any complex concepts like $h$-range problem or
Additive Number Theory. To put things in a slightly different way, the
hyper-systolic algorithm is a complex way to reconstruct combination
of rowwize ring and columwize summation on a 2D network by a sequence of
shift operations in 1-D ring network.

Thus, as far as the $C_s r$ term is small, our algorithm and the
hyper-systolic algorithm show the same scaling. However, since  $C_s r$
term  would almost always limit the scaling, the broadcast version of
our algorithm is almost always better than the hyper-systolic
algorithm. In addition, our algorithm is by far easier to understand
and implement.  

This simplicity of our algorithm makes it possible to extend our
algorithm to the individual timestep scheme and even to the Ahmad-Cohen scheme,
which will be discussed in the following sections.

\section{Application to Individual timestep}

\subsection{Standard individual timestep}

If we use the broadcast version as the base, the extension to the
individual timestep method is trivial. Instead of broadcasting all
particles in the first stage, we broadcast only the particles in the
current block. In the following steps, we always send only data
related with the particles in the current block.  Everything else is
the same as in the case of the shared timestep algorithm. Using the same
assumption of $n\sim N^{2/3}$, we have
\begin{equation}
T_{\rm ind,2D} = C_f N^{5/3}/r^2 + 2(C_cN^{2/3}/r + C_s\log_2r),
\label{eq:tind2D}
\end{equation}
and
\begin{equation}
p_{\rm half,ind,2D} \sim \cases{(NC_f/2C_c)^2 & ($N<N_{\rm c,ind,2D}$),\cr
                              N^{5/3}C_f/C_s\log_2(N^{5/3}C_f/2C_s) & ($N>N_{\rm c,ind,2D}$),
}
\label{eq:p2Dind}
\end{equation}
where $N_{\rm c,ind,2D}$ is now given by the following implicit equation
\begin{equation}
N_{\rm c,ind,2D}^{1/3}\log_2 \left(\frac{N_{\rm c,ind,2D}^{5/3}C_f}{2C_s}\right)
  = \frac{4C_c^2}{C_fC_s}.
\end{equation}

For the example values in table 1, the value of $N_{\rm c,ind,2D}$ is
fairly small. So we can use only $O(N^{5/3})$ processors. However,
this is still much larger than the number of processors that can be used
with 1D implementation of the individual timestep algorithm.

The same load-balance problem as we have discussed in the case of the 
copy algorithm occurs with this method. We need some load-balancing
strategy to actually use this method.

\subsection{Application to the Ahmad-Cohen scheme}

The difference from the individual timestep scheme is that the
neighbor list is created/used to calculate the forces. the neighbor
list for forces from particles in group $j$ to particles in group $i$
is created, stored and used only by processor $q_{ij}$. Therefore,
there is no increase in the communication cost, except for the
summation of the number of neighbors. The total calculation time and
the 50\% efficiency processor
count are given by:
\begin{equation}
T_{\rm AC,2D} = C_f N^{17/12}/r^2 + 2(C_cN^{2/3}/r + C_s\log_2r),
\label{eq:tAC2D}
\end{equation}
and
\begin{equation}
p_{\rm half,AC,2D} \sim \cases{N^{3/2}(C_f/2C_c)^2 & ($N<N_{\rm c,AC,2D}$),\cr
                              N^{17/12}C_f/C_s\log_2(N^{17/12}C_f/2C_s) & ($N>N_{\rm c,AC,2D}$),
}
\label{eq:p2DAC}
\end{equation}
where $N_{\rm c,AC,2D}$ is now given by the following implicit equation
\begin{equation}
N_{\rm c,AC,2D}^{1/12}\log_2 \left(\frac{N_{\rm c,AC,2D}^{17/12}C_f}{2C_s}\right)
  = \frac{8C_c^2}{C_fC_s}.
\end{equation}

Note that, in this  case, whether or not $N>N_{\rm c,AC,2D}$ makes very
small difference for the number of processors, since the difference is 
only of the order of $(N/N_{\rm c,AC,2D})^{1/12}$. Thus, practically we
can say that we can use $O(N^{3/2})$  processors with the 2D version
of the Ahmad-Cohen scheme.

\section{Comparison of traditional and proposed algorithms}

In this section, we present the theoretical comparison of the proposed 
algorithm and the traditional one-dimensional algorithm. First we show
the result for  the case of Myrinet-like fast network. 

\begin{figure}
\begin{center}
\leavevmode
\epsfxsize = 8 cm
\epsffile{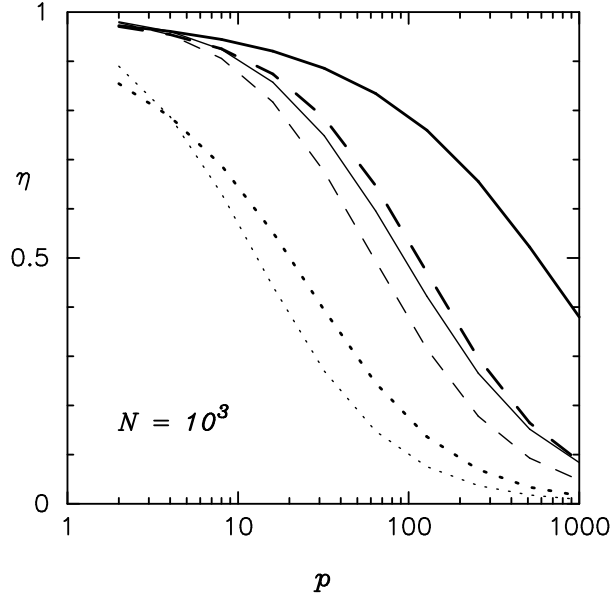}
\caption{Efficiencies of various algorithms discussed in the text. Thin 
curves correspond to 1D algorithms and thick curves to 2D
algorithms. Solid, dashed and dotted curves represent shared timestep, 
individual timestep and Ahmad-Cohen scheme, respectively. Number of
particles is $10^3$ and a Myrinet-like network is assumed.}
\label{fig:n3}
\end{center}
\end{figure}

\begin{figure}
\begin{center}
\leavevmode
\epsfxsize = 8 cm
\epsffile{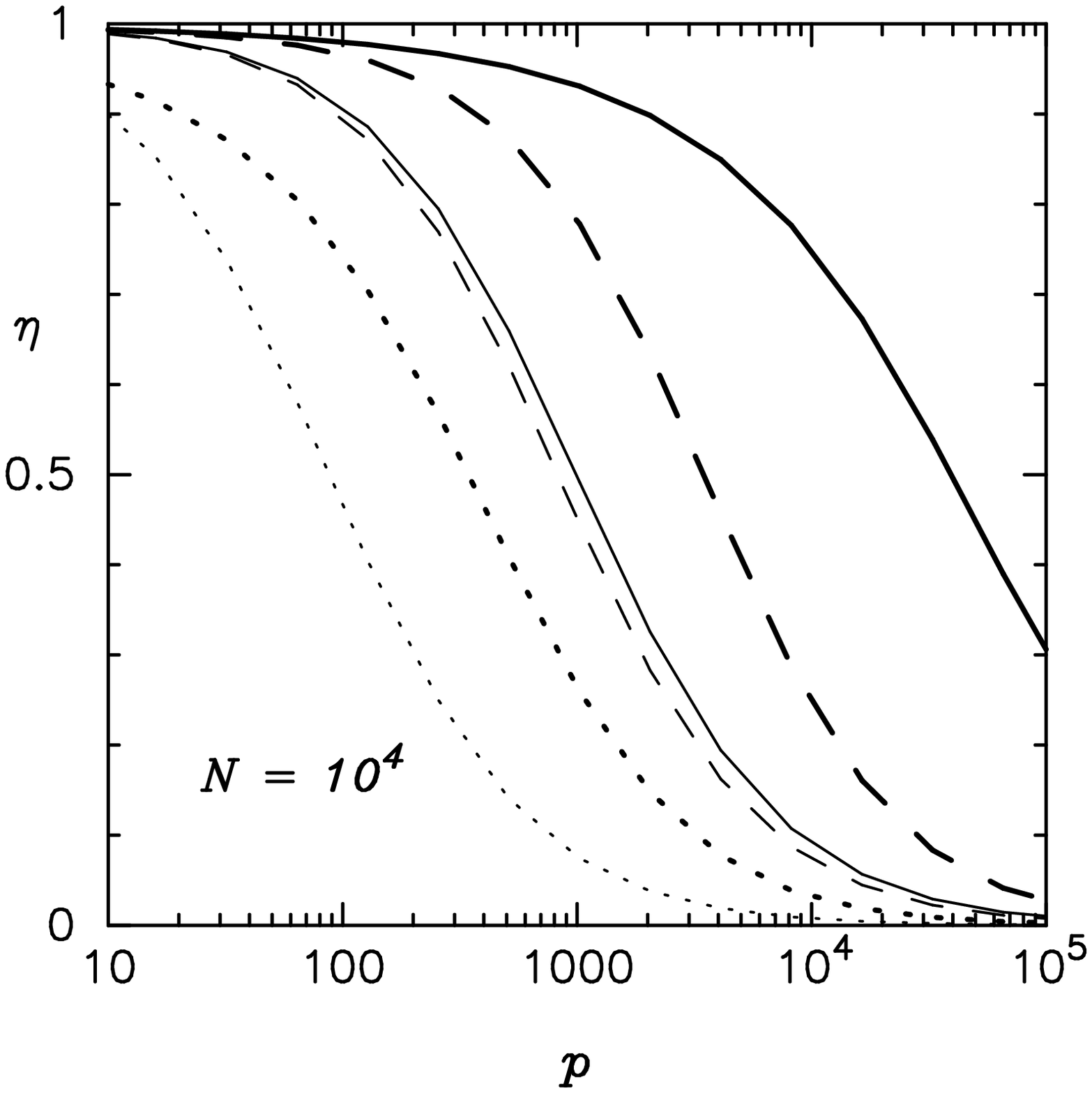}
\caption{Same as figure \protect \ref{fig:n3} but for $N=10^4$.}
\label{fig:n4}
\end{center}
\end{figure}

\begin{figure}
\begin{center}
\leavevmode
\epsfxsize = 8 cm
\epsffile{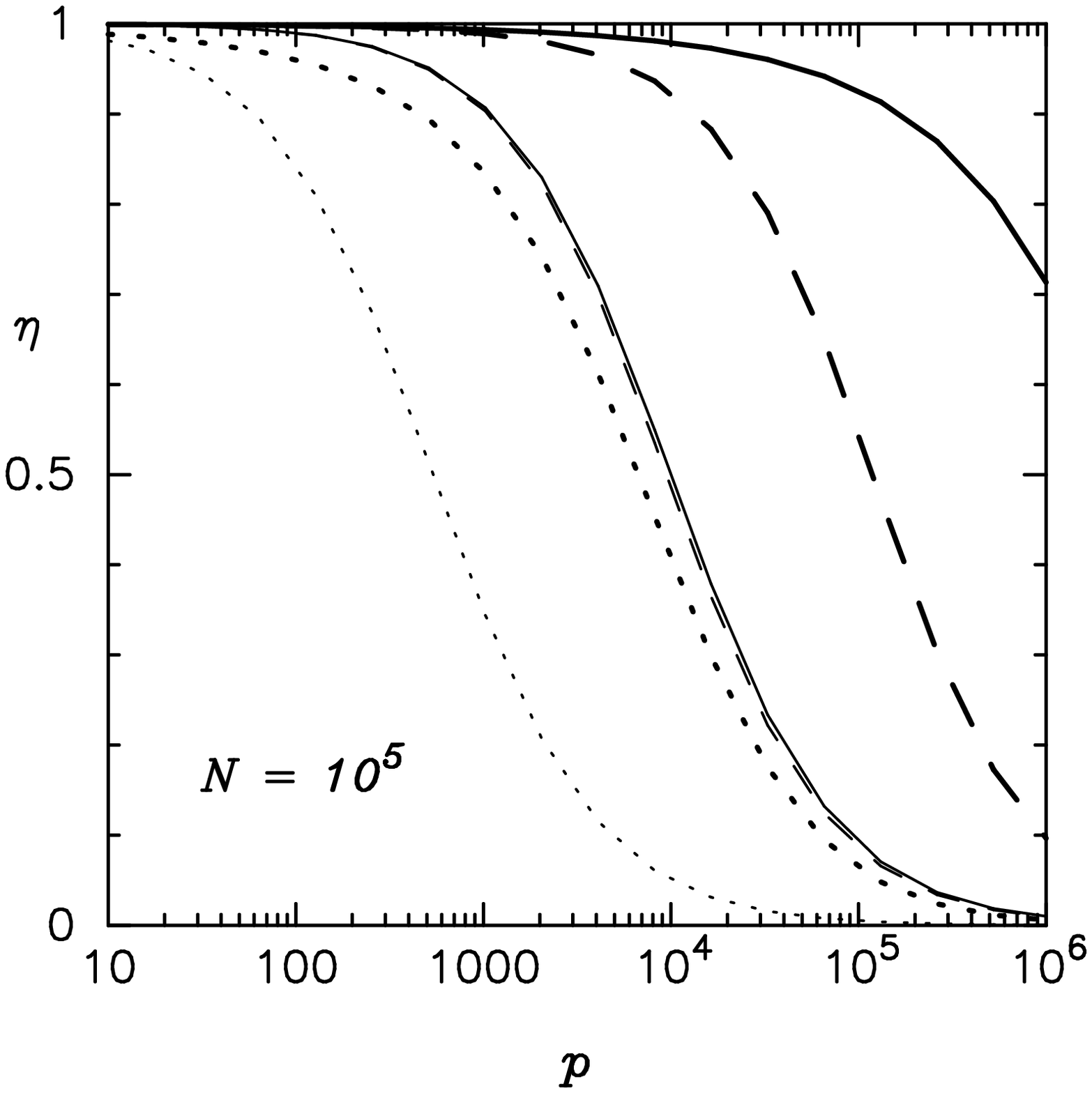}
\caption{Same as figure \protect \ref{fig:n3} but for $N=10^5$.}
\label{fig:n5}
\end{center}
\end{figure}

Figures \ref{fig:n3} to \ref{fig:n5} show the efficiencies for three
different values of $N$ as the function of the number of processors
$p$. It is clear that 2D algorithms allow us to use much larger number 
of processors compared to their 1D counterparts. The gain is larger for
larger $N$, but becomes smaller as we use more advanced
algorithms. The gain for individual timestep is smaller than that for
shared timestep, and that for the Ahmad-Cohen scheme is even smaller.

Even so, the gain in the processor count is more than a factor of 5,
for the case of the Ahmad-Cohen scheme and $N=10^4$. We believe this
is quite a large gain in the parallel efficiency. 

\begin{figure}
\begin{center}
\leavevmode
\epsfxsize = 8 cm
\epsffile{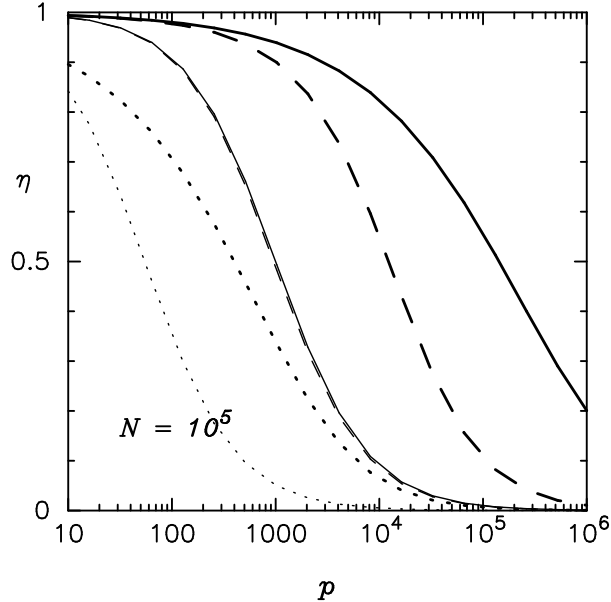}
\caption{Same as figure \protect \ref{fig:n5} but for Fast Ethernet.}
\label{fig:n5FE}
\end{center}
\end{figure}

Figure \ref{fig:n5FE} shows the efficiencies for the case of the Fast
Ethernet. With the 2D algorithm we can  use more than 500 processors
even with Ahmad Cohen scheme, for $N=10^5$. PC clusters
with inexpensive networks seem to be  very attractive platforms to
implement parallel version of the Ahmad-Cohen scheme. 

\section{Combination with GRAPE}

The only thing GRAPE does is to greatly reduce the value of
$C_f$. Thus, the same 2D network of processors each with one GRAPE
processor works fine, if the cost of the frontend is less than that of 
a GRAPE processor.

GRAPE-6 achieves
essentially the same effect as this 2D processor grid, but using only
$r$ hosts and $r^2$ GRAPE processors connected with a rather elaborate
multistage networks. In hindsight, such an elaborate network is
unnecessary, if the fast network is available for a low cost. 

From the point of view of the scaling relations, what a GRAPE hardware 
changes is simply 
$C_f$. If we attach a 1Tflops GRAPE hardware to a 1 Gflops host, we
reduce $C_f$ by a factor of $10^3$. This means that the limiting
factor for the number of processors is almost always $C_c$ and not
$C_s$.  Thus, for a parallel GRAPE system, high-throughput,
high-latency network such as Gigabit Ethernet is a practical choice.

\begin{figure}
\begin{center}
\leavevmode
\epsfxsize = 8 cm
\epsffile{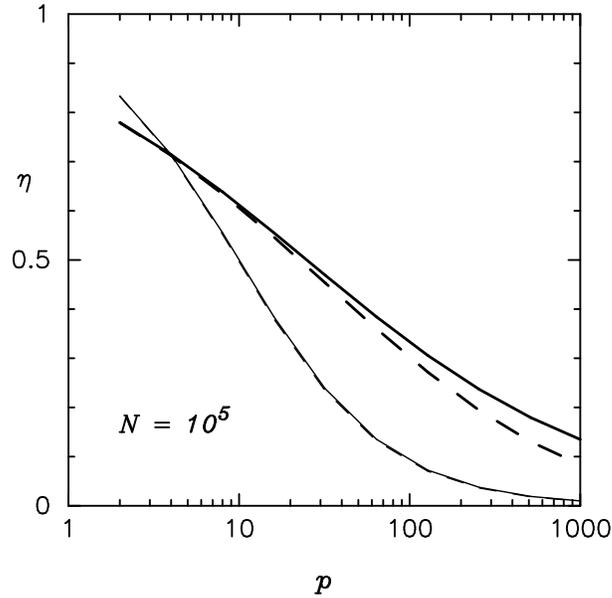}
\caption{Same as figure \protect \ref{fig:n5} but for $C_f=3\times
10^{-11}$. }
\label{fig:n5G6}
\end{center}
\end{figure}

\begin{figure}
\begin{center}
\leavevmode
\epsfxsize = 8 cm
\epsffile{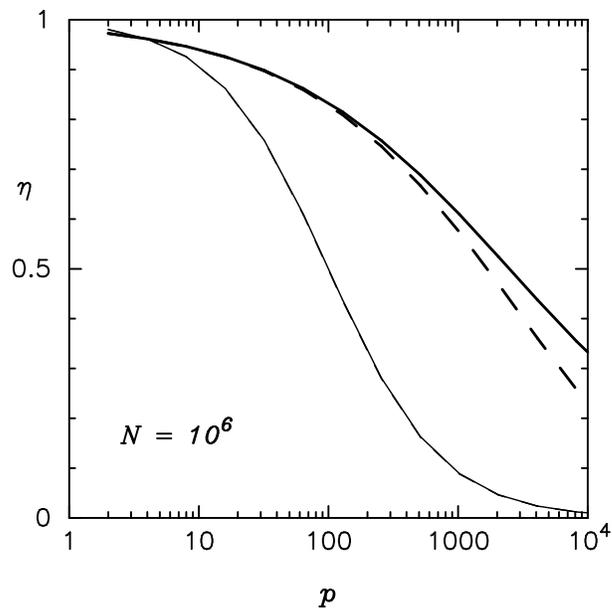}
\caption{Same as figure \protect \ref{fig:n5G6} but for $N=10^{6}$. }
\label{fig:n6G6}
\end{center}
\end{figure}

Figures \ref{fig:n5G6} and \ref{fig:n6G6} shows the efficiencies for
GRAPE-6 system. Since $C_f$ is much smaller, the number of processors
we can use becomes much smaller. Of course, in the case of 1D
algorithms, the actual speed we can achieve does not depend on $C_f$,
since the total speed is $p/C_f$ and $p_{\rm half}$ is proportional to
$C_f$. Figure \ref{fig:n6G6} indicates that we can achieve the speed
of multiple petaflops with currently available technology, by
configuring several thousand GRAPE-6 boards into a 2D network.

\section{Summary}

We described a new two-dimensional algorithm to
implement the direct summation method on distributed-memory parallel
computers. The basic idea of the new algorithm is to organize
processors to $r\times r$ 2D network, and let the data be shared both
rowwize and columnwize. In this way, we can reduce the communication
cost from $O(N)$ of the previously know algorithms to $O(N/r)$.
 	
For the case of the shared timestep algorithm, the new algorithm
behaves in essentially  the same as the ``regular bases'' version of
the hyper-systolic algorithm does. However, with the broadcast version
of our algorithm the communication overhead is reduced, which resulted
in the better scaling. Also, our algorithm is much simpler, which
helped us to extend our algorithm to individual timestep and the Ahmad-Cohen
scheme, as well as combination with GRAPE hardwares.

For all cases, compared to the previously known algorithm, the number
of processors we can use without losing efficiency is almost {\it
squared}. This is a quite large improvement in the efficiency of a parallel
algorithm.

Usually, a paper which proposes a new parallel algorithm should offer
the verification of the concept, by means of the timing measurement of
actual implementation.  In this paper we omit this verification,
because we believe it's important to let those who working on
hyper-systolic algorithms be aware of simpler alternatives.

In this paper, we assumed a network with full connectivity. This
assumption is okay with small PC clusters, but not true on large MPPs.
In this case, parallel efficiency of 1D algorithm is significantly
reduced, and relative gain of 2D algorithm would become much larger. 

\section*{Acknowledgments}

I thank Rainer Spurzem and Yoko Funato for valuable
discussions.  This work is supported in part by the Research for the
Future Program of Japan Society for the Promotion of Science
(JSPS-RFTP97P01102).

\def\araa{Annual Review of Astronomy and Astrophysics }
\def\aap{Astronomy and Astrophysics }
\def\aj{The Astronomical Journal }
\def\apj{The Astrophysical Journal }
\def\APJ{The Astrophysical Journal }
\def\apjl{The Astrophysical Journal Letters }
\def\apjs{The Astrophysical Journal Supplement Series }
\def\apss{Astrophysics and Space Science }
\def\ajl{The Astronomical Journal }
\def\pasj{Publications of the Astronomical Society of Japan }
\def\mn{Monthly Notices of Royal Astronomical Society }
\def\MN{Monthly Notices of Royal Astronomical Society  }
\def\mnras{Monthly Notices of Royal Astronomical Society }
\def\nat{Nature }
\def\jcp{Journal of Computational Physics }


\end{document}